\documentclass[english,twocolumn,aps,prl, superscriptaddress,showpacs, amsmath,floatfix,,nofootinbib]{revtex4-2}

\usepackage{chemformula} 
\usepackage[T1]{fontenc} 
\usepackage{color,xcolor}
\usepackage{verbatim}

\usepackage{natbib}
\usepackage[T1]{fontenc}
\usepackage{units}
\usepackage{amssymb,amsmath}
\usepackage{graphicx}
\usepackage{esint}
\usepackage{bm}
\usepackage{ulem}
\usepackage[colorlinks=true, citecolor=blue]{hyperref}
\setcitestyle{journalcolor=blue}

\begin{document}

\title
  {Energy Gap Modulation in Proximitized Superconducting Puddles of Graphene }

\author{Yuxiao Wu*}
\affiliation{Department of Physics and Jack and Pearl Resnick Institute and Institute of Nanotechnology and Advanced Materials,
  Bar-Ilan University, Ramat-Gan 52900, Israel}
\author{Udit Khanna*}
\affiliation{Department of Physics and Jack and Pearl Resnick Institute and Institute of Nanotechnology and Advanced Materials,
  Bar-Ilan University, Ramat-Gan 52900, Israel}
\affiliation{Theoretical Physics Division, Physical Research Laboratory, Navrangpura, Ahmedabad-380009, India}
\author{Eyal Walach*}
\affiliation{Department of Physics and Jack and Pearl Resnick Institute and Institute of Nanotechnology and Advanced Materials,
  Bar-Ilan University, Ramat-Gan 52900, Israel}

\author{Efrat Shimshoni}
\affiliation{Department of Physics and Jack and Pearl Resnick Institute and Institute of Nanotechnology and Advanced Materials,
  Bar-Ilan University, Ramat-Gan 52900, Israel}

\author{Aviad Frydman}
\affiliation{Department of Physics and Jack and Pearl Resnick Institute and Institute of Nanotechnology and Advanced Materials,
  Bar-Ilan University, Ramat-Gan 52900, Israel}
\email{aviad.frydman@gmail.com}


\date{\today}

\begin{abstract}
We investigated proximity-induced superconductivity in a graphene–insulating InO bilayer system through gate-controlled transport measurements. Distinct oscillations in the differential conductance are observed across both the electron and hole doping regimes, with oscillation amplitudes increasing as the chemical potential moves away from the Dirac point. These findings are explained using a theoretical model of a normal-superconductor-normal (NSN) junction, which addresses reflection and transmission probabilities at normal incidence. From this model, we extract key parameters for the proximitized graphene, including the superconducting energy gap $\Delta$ and the effective length scale $L_s$ of the superconducting regions. Near the Dirac point, we observe a minimal $L_s$ and a maximal $\Delta$, aligning with the theory that the gap in strongly disordered superconductors increases as the coherence length of localized pairs decreases. This suggests that spatial confinement in a low-density superconductor leads to an effective increase in the superconducting gap. 

* equal contribution

\end{abstract}

\maketitle

\section{Introduction}

The superconducting proximity effect \cite{PhysRev.117.672} describes the induction of superconductivity in a normal conductor through a normal metal/superconducting (NS) interface. Typically, the superconducting order parameter decays exponentially on the normal side over a finite normal-state coherence length $\xi_n$. Consequently, when a conventional superconductor is in contact with monolayer graphene, a two-dimensional (2D) semimetal composed of a single atomic layer of carbon, Cooper pairs enter the monolayer, opening a 
superconducting energy gap in the band structure of graphene. Inducing superconductivity into graphene and other topological materials in quantum Hall states offers promising pathways for studying Majorana Fermions, topological superconductivity \cite{PhysRevLett.100.096407,PhysRevLett.105.077001,Alicea_2012,Das2012-nf,RevModPhys.83.1057,Hart2014-sv} and realizing topological quantum computing \cite{KITAEV20032,KITAEV20062,AndrewSteane1998,RevModPhys.80.1083}. Additionally, the properties of proximitized graphene, particularly its enhancement of superconductivity \cite{PhysRevB.101.054503,Lee_2018} have attracted significant interest from condensed matter physicists. The critical temperature and current\cite{PhysRevLett.104.047001,Allain2012-bn,Han2014-du,PhysRevB.105.L100507} of the graphene-superconductor heterostructure have been found to be tunable with respect to the charge carrier density. Moreover, other exotic phenomena such as crossed Andreev reflections \cite{Lee2017-qy,PhysRevX.12.021057} and quantum interference \cite{Heersche2007-kk, Ben_Shalom2016-rn,Huang2022-or} have been observed in such systems.

While substantial research has been conducted on bilayer graphene and conventional superconductors under extreme conditions to enhance their properties and explore potential applications, the proximity coupling between graphene and 2D disordered superconductors remains poorly understood. This is particularly true in the context of the disorder-driven superconductor-insulator transition (SIT) \cite{dubi2007nature,baturina2007localized,sherman2014effect,sacepe2008disorder}, where emergent granular superconductivity has been observed. In this transition, the interplay between the electronic charging energy of the grains and the Josephson coupling between them plays a crucial role in its appearance. 

Amorphous indium oxide (InO) films \cite{KOWAL1994783} exemplify such 2D disordered superconductors. Despite being morphologically uniform, InO is found to exhibit emergent granular superconductivity embedded in an insulating matrix. Remarkably, even in the highly insulating state, signs for residual superconductivity were observed experimentally such as giant magnetoresistance \cite{PhysRevLett.92.107005}, Little-Parks oscillations \cite{PhysRevLett.109.167002}, and the persistence of a superconducting energy gap \cite{PhysRevLett.108.177006}, consistent with theoretical works that predicted the emergence of superconducting grains in the insulating matrix \cite{Bouadim2011-ny}.  This exotic state has been dubbed a "Bosonic insulator". 

A unique feature of an InO film is its low carrier density $(n = 10^{19} - 10^{20} \text{cm}^{-3})$ \cite{ZOvadyahu_1986}, nearly two orders of magnitude lower than in typical metals.  When a monolayer of graphene is brought into contact with InO, two key effects arise. First, the low electronic doping from InO allows for external gating to tune the system from an electron-doped superconductor to a hole-doped superconductor through the charge neutrality point (CNP), potentially realizing a strongly coupled Bose-Einstein condensate (BEC) superconductor \cite{RevModPhys.96.025002}. Second, the electrical conductivity of graphene itself is tunable via gating, enabling it to function as a buffer layer that modulates the Josephson coupling between superconducting puddles induced by InO. This tunability makes the InO/graphene bilayer a promising platform for exploring the SIT through electrostatic gating.

\begin{figure*} [ht]
\includegraphics[width=1\textwidth]{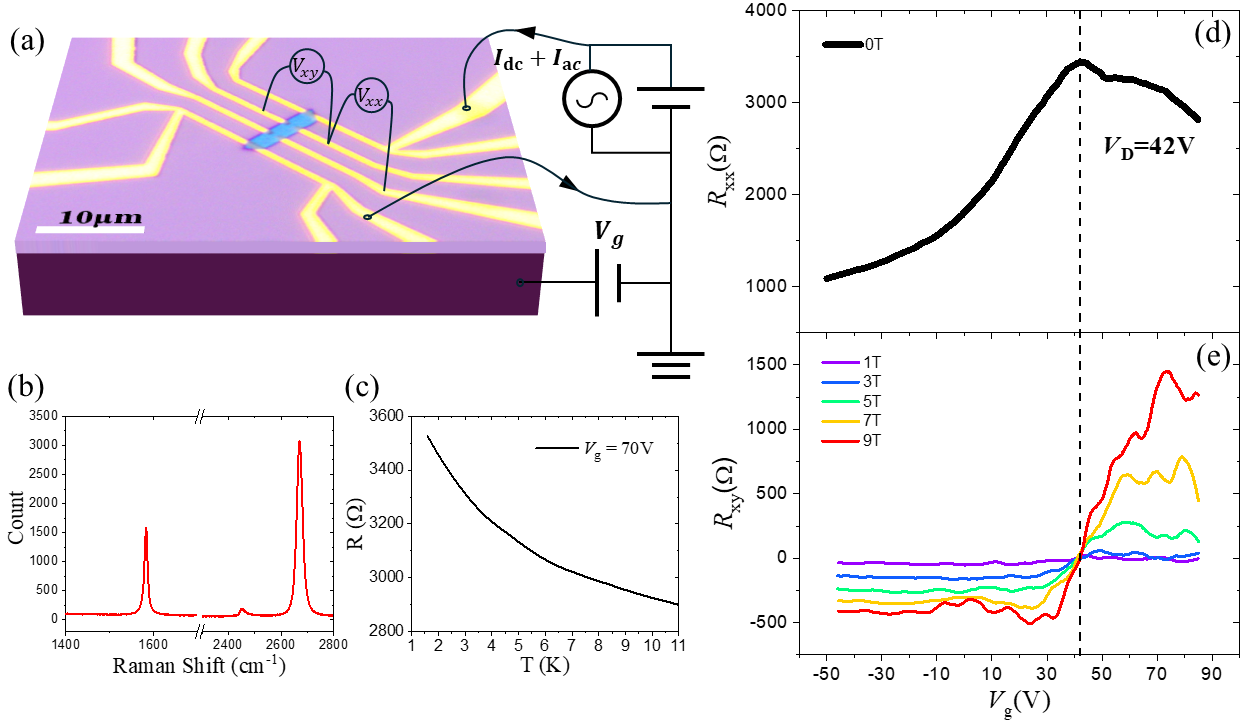} 
\caption{
(a) Optical image of the device and measurement setup configured for gated transport measurements and differential conductance measurement, with a source current, $I_{dc} + I_{ac}$, applied across the device, and voltage probes for longitudinal ($ V_{xx} $) and Hall ($ V_{xy}$) resistances. (b) Raman spectrum identifying the presence of monolayer graphene after exfoliation. (c) Resistance versus temperature plot at a fixed gate voltage $V_g = 70$V. The increase in resistance with lowering temperature indicates the insulating feature of the sample.  (d) longitudinal resistance ($ R_{xx} $) and (e) Hall resistance ($ R_{xy} $) as a function of gate voltage $V_g$ at $T=1.6$K, under different perpendicular magnetic fields (up to 9T). The curves reveal quantum oscillations, which are the results of the superconducting fluctuations for low magnetic fields and quantum Hall effect at high magnetic fields. The Dirac point is indicated by a dashed line, based on the fact that $ R_{xx} $ peaks at $V_g = 42$V, where the carrier density in graphene is minimized and the global charge neutrality revealed by the crossing point of curves in (e) where $ R_{xy} = 0 $ . 
}

\label{fig1}
\end{figure*}

In this work, we present a combined experimental and theoretical study on a bilayer of graphene and insulating-phase InO. Our experiments focus on differential conductance of the bilayer across a broad doping range around the Dirac point, controlled via back gating. We develop a theoretical model to fit the results and extract two key parameters: $L_s$, the effective length scale of the superconducting puddles, and $\Delta$, the superconducting energy gap. The main findings are as follows: (1) conductance oscillations appear in the differential conductance spectra in both electron and hole doping regimes, with enhanced oscillation amplitude when the chemical potential deviates from the Dirac point; (2) our analysis indicates that superconducting puddles shrink near the Dirac point while the superconducting energy gap, $\Delta$, increases. This is consistent with the existing theory of emergent granularity \cite{Bouadim2011-ny} of disordered superconductors, which shows that $\Delta$ increases with increasing disorder since the superconducting puddle size shrinks.

\begin{figure*}[t]
\includegraphics[width=0.8 \textwidth]{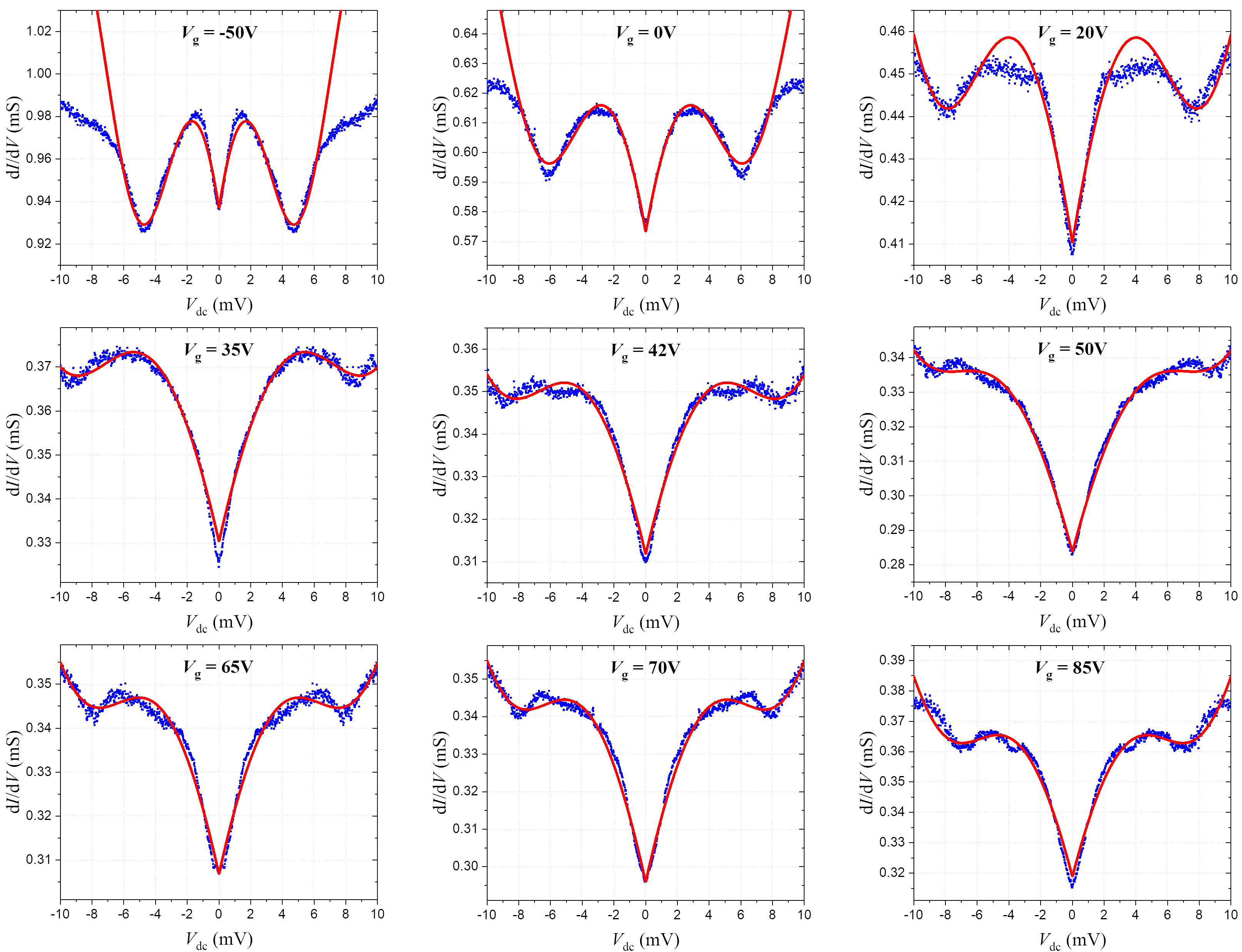} 
\caption{
Differential conductance ($\text{d}I/\text{d}V$) measured at 1.6K and 0T as a function of dc bias voltage at various gate voltages. Oscillations are clearly visible in the conductance curves, with periodic modulations that vary in amplitude as the gate voltage changes. These oscillations reflect the details of superconductivity induced by the proximity effect. The red curves represent theoretical fits taken at normal incidence, capturing the oscillatory behavior and allowing for extraction of key parameters $L_s$ and $\Delta$.
}
\label{Ndidv}
\end{figure*}

\section{Experimental results}

The samples in this work were fabricated as follows: we began with exfoliation of high quality single layer graphene (SLG) from commercial highly oriented pyrolytic graphite (HOPG) onto a $Si$ substrate with a 285nm thermally formed oxide on top. SLG was then identified by optical contrast and Raman spectrum (see fig. \ref{fig1} (c)). The substrate is then subjected to a furnace at $350^\circ C$ for high temperature annealing in a $Ar/H_2$ environment for 4-5 hours. Electrodes made of 5nm-thick Cr and 30nm-thick Au are patterned by e-beam and thermal evaporation, respectively. A 30nm-thick thin film of insulating InO is then e-beam evaporated with a partial oxygen pressure of $2.5 * 10^{-5}$ Torr such that $R_{\square}$ of the InO film is expected to be greater than $1\text{M}\Omega >> R_\text{SLG} \sim  \text{k}\Omega$s  at low temperatures, the transport in the SLG is dominating the transport properties of the bilayer. The real image of the sample is shown in Fig. \ref{fig1} (a). The sample is transferred into a He-4 cryostat with a base temperature of 1.5K and external magnetic field up to 9T.

The gate voltage dependent longitudinal resistance $R_{xx}$ and Hall resistance $R_{xy}$ at $T$=1.6K are shown in Fig.\ref{fig1}(d) and (e). The Dirac point of this device is obtained at gate voltage $V_g = 42\text{V}$, which is the global CNP identified by the crossing point in the $R_{xy}$ together with the gate voltage where the longitudinal resistance $R_{xx}$ peaks at 0T. Fluctuations are developed in both $R_{xx}$ and $R_{xy}$ as a result of universal voltage fluctuations of disordered superconductors \cite{PhysRevLett.125.147002} and Shubnikov–de Haas oscillations approaching quantum Hall states \cite{Zhang2005-pm,Novoselov2005-rd, PhysRevB.86.115412} .

We adopt a standard setup for $dI/dV$ measurements  shown in Fig.\ref{fig1}(b).
Fig. \ref{Ndidv} presents the differential conductance measured at 1.6K for different gate voltages spanning the hole ($V_g < 42$V) and electron ($V_g > 42$V) doped regime. All results are normalized at DC bias voltage $V_{dc}=10$mV. A typical zero-bias dip (ZBD) structure is found in the middle of every dI/dV curve, rendering the linear dispersion of graphene. On top of this background, prominent oscillations appear on both wings of the spectra at higher DC bias. This unique feature cannot be explained by the Blonder-Tinkham-Klapwijk (BTK) Formalism \cite{PhysRevB.25.4515} with a quasi-particle lifetime broadening $\Gamma$ \cite{PhysRevLett.41.1509} and indicates the special role of the proximitized disordered superconductivity.

\begin{figure} [b]
\includegraphics[width=0.5\textwidth]{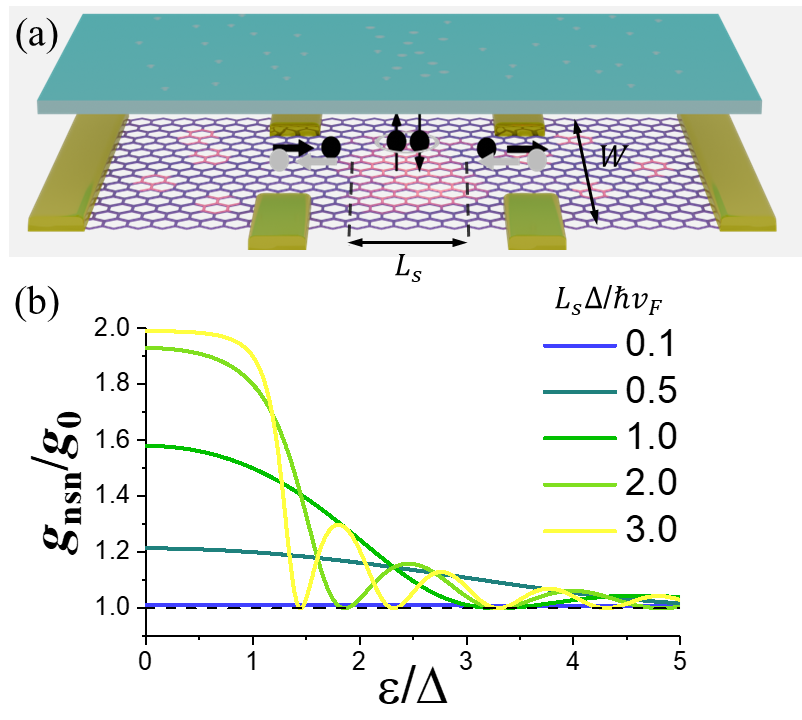}
\caption{(a) Schematic illustration of a graphene/InO bilayer. Superconducting puddles (red) and Cooper pairs are induced into the monolayer graphene, embedded in normal graphene regions (purple). Incoming/outgoing electrons (black spheres)/ holes (white spheres) undergo Andreev reflections at the NS boundary. The length of the superconducting puddle between the voltage leads is defined as $L_s$ and the sample width is denoted as $W$.(b) Conductivity of graphene NSN junction $g_{nsn}$ modulated by the conductivity of normal graphene as a function of $ \epsilon/\Delta$ at several given value of $L_s \Delta / \hbar v_F$. Conductance oscillations are found where $\epsilon > \Delta$ and their amplitude and frequency increases with increasing $L_s \Delta$.
}
\label{model}
\end{figure}

\section{Theoretical model}

Considering that an insulating InO film contains phase-incoherent superconducting puddles within an insulating ground state, it seems appropriate to interpret our sample as a normal-superconductor-normal (NSN) junction. Thus, the conductance of proximitized graphene is highly influenced by the processes of normal reflections and Andreev reflections occurring at the interfaces of the superconducting puddles. A schematic diagram of the sample is displayed in Fig. \ref{model}(a).
We define $W$ as the width of the device and $L_s$ as the length of the superconducting segment. By combining the Landauer formalism \cite{PhysRevB.41.7906} and BTK model, we find the angle-resolved differential conductance of such a NSN junction to be

\begin{equation}
    g_{NSN}\left(\alpha,\varepsilon\right) = g_Q \left(1 - \lvert r_{ee}\left(\alpha,\varepsilon\right) \rvert^2 + \lvert r_{eh}\left(\alpha,\varepsilon\right) \rvert^2\right) \cos(\alpha),
\label{conductance}
\end{equation}
where $g_Q = \frac{4e^2}{h}$ is the conductance quantum for a single channel in normal graphene,  $\alpha$ and $\varepsilon$ are respectively the incidence angle and kinetic energy of incident electrons or holes, $r_{ee}$ and $r_{eh}$ are the probability amplitudes of normal reflections and Andreev reflections, respectively, and $E_F$ is the Fermi energy of the normal states in graphene.
While BTK, considering a planar junction with $W\gg\xi$, write the overall conductance as an integral over $g(\alpha)$ with all angles contributing equally, we expect the system to be dominated by the low angles of incidence, and specifically normal incidence with $\alpha=0$ 
as detailed below.

Andreev reflections \cite{PhysRevLett.97.067007,RevModPhys.80.1337,Efetov2016-qj} in graphene have been intensively studied over the last two decades. It has been found that Andreev reflections occurring at the graphene-superconductor interface can take two distinct forms due to graphene's linear dispersion. One is Andreev retroreflection, and the other is known as specular Andreev reflection; for our parameters, $\varepsilon<\left|E_F\right|$ and we get retroreflection. Unlike regular reflection, Andreev reflections enhance the conductivity, as an electron-hole pair with momenta $\mathbf{k}$ and $-\mathbf{k}$ is created at the boundary, the opposite-charged hole is returning and two electrons are transmitted.

The probabilities of Andreev reflection, as well as regular reflection, depend strongly on the value of the incidence angle $\alpha$.
Specifically in graphene interfaces, the dominance of $\alpha\approx 0$ is greatly enhanced due to
Klein tunneling \cite{Katsnelson2006-td,RevModPhys.80.1337,Perconte2018-gi} which occurs when a relativistic particle normally collides with an energy barrier. It is known that the reflection probability $R = 0$ and transmission probability $T = 1$ when the incidence angle $\alpha=0$, rendering the energy barrier effectively transparent. In the case of Andreev reflections, $R$ is still $0$, but $T$ and $R_A$ may vary. Either way, the conductivity is enhanced for the case of $\alpha=0$.

When discussing a small, disordered puddle of width $W\approx\xi$, the possible incidence angles are discrete rather than continuous. The number of "modes", which in the continuum limit can be taken to be $\frac{W(E_F+\varepsilon)}{\pi\hbar v_F}$, is now discrete and the allowed angles depend on $W$. Furthermore, as can be seen in the Supplemental Material, $r_{ee}$ and $r_{eh}$ oscillate as a function of $k_x \sim \cos\alpha$, which is close to constant for $\alpha\approx0$  (as $\frac{\partial k_x}{\partial \alpha}\mid_{\alpha=0} = 0$) but changes more and more rapidly as $\alpha$ increases.
As $W$ is disordered, the contributions of most modes are averaged out 
except for the $\alpha=0$ mode, which persists independently of $W$.

In view of the above, it seems reasonable to assume that the conduction of the NSN junction is dominated by the single channel corresponding to normal incidence. In such case we compute $r_{ee}$ and $r_{eh}$ by imposing continuity of the wavefunctions at the two NS junctions and we find $r_{ee}=0$ (due to Klein tunneling) and
\begin{equation}
 r_{eh}=-i\frac{\text{sin}(\frac{L_s\Delta}{\hbar v_F} \text{sinh}\beta)  }{\text{cos}(\frac{L_s\Delta}{\hbar v_F}\text{sinh}\beta)\text{sinh}\beta  -i\text{sin}(\frac{L_s\Delta}{\hbar v_F}\text{sinh}\beta)\text{cosh}\beta  }
\end{equation}
where $\beta = i \text{cos}^{-1}(\epsilon/\Delta)$ for $\epsilon<\Delta$ and $\beta = \text{cosh}^{-1}(\epsilon/\Delta)$ for $\epsilon > \Delta$, and $v_F\approx 10^6 \text{m/s} $ is the Fermi velocity of electrons in graphene.

We find that the probability of Andreev reflection depends entirely on $ L_s \Delta $ and $ \epsilon/\Delta $. 
Thus, by employing Eq. (\ref{conductance}), and taking into account that the effective number of channels is proportional to the density of states of graphene $\propto (E_F+\varepsilon)$, 
the final form of the NSN junction conductance is written as

\begin{equation}
    g_{NSN} = C (E_F+\epsilon) \times (1+\frac{1-\text{cos}(\frac{2L_s\Delta}{\hbar v_F} \text{sinh}\beta)  }{\text{cosh}(2\beta)-\text{cos}(\frac{2 L_s\Delta}{\hbar v_F}\text{sinh}\beta  )  })
\label{gnsn}
\end{equation}
where $C$ is a normalization factor. Clearly in agreement with the experimental results, the conductance oscillates with increasing $\beta$ (or kinetic energy $\epsilon$). The conductance enhancement and oscillations for different values of $L_s\Delta$ due to Andreev reflections are plotted in Fig. \ref{model} (b). Theoretically we expect the conductance to reach a minimum for all $\beta$ values that satisfy $L_s\Delta \text{sinh}\beta = n \pi$ for $n = 1,2,3...$, indicating the energy levels where the barrier is transparent for normally incident electrons and $r_{eh} = r_{ee} = 0$.
In this case, since the probabilities of both normal reflections and Andreev reflections vanish for electrons or holes at these levels, the conductance of the NSN junction is dominant by the elastic cotunneling process \cite{PhysRevLett.95.229901}.

\section{Discussion}

Fig. \ref{Ndidv} depicts the $\mathrm{d}I/\mathrm{d}V$ curves measured at different gate voltages together with a fit to the analytic solution of Eq. (\ref{gnsn}). It is seen that the results fit the theory  well in the vicinity of the Dirac point including the clear observed conductance oscillations. On the electronic side at high $V_{dc}$ the results deviate from the theoretical curve presumably due to disorder and strong electron-electron interactions at high doping and bias voltage.

\begin{figure} [ht]
\includegraphics[width=0.4\textwidth]{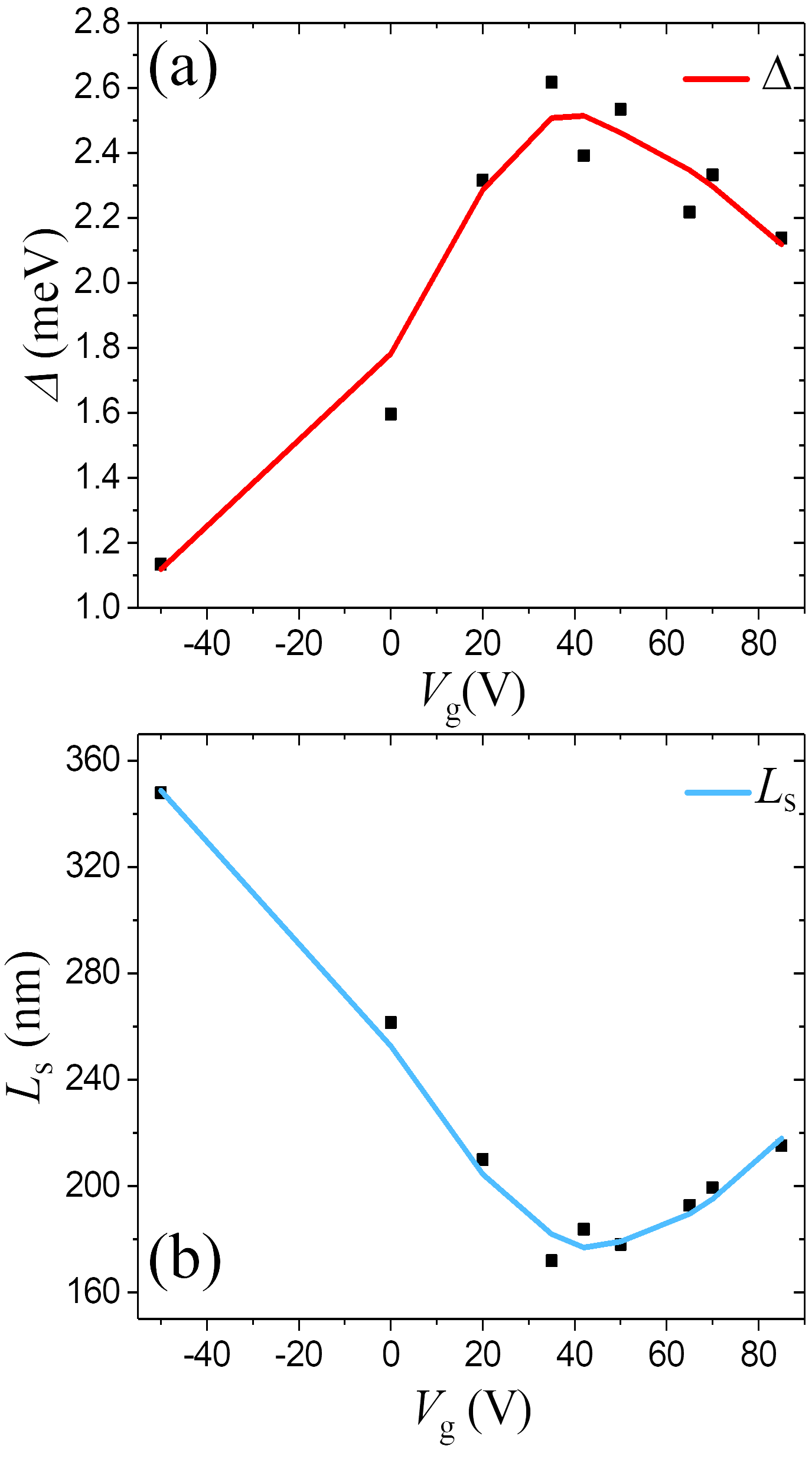}
\caption{
Extracted gate dependence of the (a) superconducting energy gap $\Delta$ (red) and (b) effective length $L_s$ (blue) of the superconducting segment in the sample. As the gate voltage increases, both curves show non-monotonic behaviors, suggesting that the gate voltage significantly influences the superconducting properties by modulating the proximity effect within the heterostructure. Generally, $\Delta$ exhibits an inverse relationship with the $L_s$ and, notably, $\Delta$ exhibits a maximum around the Dirac point $V_g = 42$V while $L_s$ reaches a minimum.
}
\label{Delta&Ls}
\end{figure}

From the fits, we obtained key parameters, $L_s$ and $\Delta$, and plotted them as functions of gate voltage in Fig. \ref{Delta&Ls} (a) and (b). It is seen that $L_s$ varies between 180 nm and 350 nm, reaching a minimum near the Dirac point. On the other hand, $\Delta$ ranges from 1.1 meV to 2.5 meV, reaching a maximum near the Dirac point. This indicates that application of a gate increases the size of the superconducting puddles in the sample, and, at the same time, decreases the superconducting energy pap within the puddles

We recall that theoretical results from the attractive Hubbard model on the insulating side of the superconductor-insulator transition \cite{Bouadim2011-ny} suggest that, strong disorder causes an increase of the superconducting energy gap due to confinement of the superconducting regions into a small localization volume, $\xi^2_{loc}$ which enhances the attraction between paired electrons. Our observations are consistent with this picture, correlating that increase of spatial confinement with gap enhancement. However, in our case it is not the disorder which is varied but the tuning of the normal graphene chemical potential close to the Dirac point. This reduces the available density of states, allowing smaller regions to acquire superconductivity from the InO. Consequently, by controlling the confinement of the superconducting region, we have achieved an increase in superconducting energy gap of almost 150\%.

Although our theoretical model successfully explains the experimental results, the theory has certain caveats when computing the conductivity of graphene NSN junctions. First, this theory only works for a quasi-one-dimensional ballistic graphene NSN junction without taking into account the spatial distribution of the superconducting puddles. Second, possible corrections in the conductance of graphene itself due to effects like electron-electron interactions and weak localization \cite{Somphonsane2020} are not included in this model. It also remains unclear why a simple model that considers only normal incidence is sufficient. While the angle dependence of Andreev reflection provides some insight, we speculate that the sample geometry also plays a significant role. Unlike a continuum model that integrates over all angles to determine the conductivity of a graphene NS junction \cite{PhysRevLett.97.067007}, the superconducting puddle in our bilayer has a confined width relative to the overall sample width and may have an ellipsoid-like geometry, which naturally limits the range of angles contributing to transport in the system.


In summary, our study demonstrates the occurrence of conductance oscillations in the differential conductance spectra of a graphene/insulating InO bilayer system, offering insights into proximitized superconductivity under gate modulation. By controlling the gate voltage, we observed variations in the effective superconducting segment length $L_s$, and the superconducting energy gap $\Delta$, with both quantities showing a distinct dependence on the proximity to the Dirac point. Near the Dirac point, superconducting puddles reach their smallest size while the superconducting gap attains its peak value. This behavior aligns with theoretical models that suggest enhanced energy gap due to spatial confinement under increasing disorder, though here achieved via gate control without adjusting disorder. Thus, our findings underscore the potential of gate-tunable proximitized systems in superconducting energy gap engineering and studying quantum tunneling in complex superconducting systems, particularly within low-density disordered superconductors. Further studies on varied disorder profiles may expand on these findings and probe deeper into the mechanisms of gap enhancement and quantum interference in disordered superconducting systems.

\begin{acknowledgements}
  The authors thank I. Volotsenko, J.F. Stein, M. Ismail, N. Shabi and G.N. Daptary for technical help and N. Trivedi for useful discussions. Y.W. and A.F were supported by the Israel National Fund, ISF grant number 1499/21. U.K., E.W. and E.S. acknowledge the support of the Israel Science Foundation (ISF) Grant No. 993/19.
  
\end{acknowledgements}

\bibliography{Bibliography}

\clearpage
\appendix
\onecolumngrid
\section*{Supplemental Material}
\addcontentsline{toc}{section}{Supplemental Material}

\noindent
In this supplementary material, we detail the methodology used to address the ballistic conductance across a superconducting region of finite length in graphene. We consider a planar geometry in which the two NS interfaces are aligned parallel to the $y$-direction. The length of the superconducting segment along the $x$-axis is denoted as $L_{s}$. We derive the formula for the conductance of a graphene NSN junction within a general 2D planar model, beginning by expressing the eigenstates in both the normal and superconducting regions of graphene. Next, we describe the process of calculating the reflection and transmission amplitudes by enforcing the continuity of the wavefunctions at the two NS junctions. These amplitudes are then used to determine the conductance of the system. In the final section, we apply the general model to the case of normal incidence ($q=0$) and derive an analytical expression for the conductance, which is in good agreement with the experimental results presented in the main text. (Throughout this note, we set $\hbar v_{F} = 1$.)

\subsection*{States in the Normal Regions}

\noindent
The normal (unproximitized) region in graphene may be described by the Hamiltonian
\begin{align}
  H_N = \left( \begin{array}{cc}
    H_0 - E_F & 0 \\
    0 & E_F - H_0 \end{array} \right), 
\end{align}
where $H_0 = -i (\sigma_x \partial_x + \sigma_y \partial_y)$, $E_F$ is the Fermi energy and $\sigma$ represents 
the sublattice degree of freedom. 
We have employed the BdG basis so that the upper (lower) block of $H_{N}$ corresponds to 
electrons (holes) that belong to opposite valleys. There is an additional degeneracy of 4 due to spin and valley.\\

\noindent
For each wavevector $\vec{p}$, $H_{N}$ has 4 eigenvalues: $\epsilon_{e \pm} = -E_{F} \pm \mathfrak{p}$ and 
$\epsilon_{h \pm} = E_{F} \pm \mathfrak{p}$ where $\mathfrak{p} = \sqrt{p_{x}^{2} + q^{2}} > 0$.  
The corresponding eigenfunctions are, 
\begin{align}
  &\psi_{e+} (\epsilon, \vec{r}\,) = e^{i \vec{p}_{e} \cdot \vec{r}} 
  \left( \begin{array}{c} \frac{p_{x e} - i q}{\mathfrak{p}_e} \\ 1 \\ 0 \\ 0 \end{array} \right) 
    \, \, \text{, } \,\,
  \psi_{e-} (\epsilon, \vec{r}\,) = e^{i \vec{p}_{e} \cdot \vec{r}} 
  \left( \begin{array}{c} -\frac{p_{x e} - i q}{\mathfrak{p}_e} \\ 1 \\ 0 \\ 0 \end{array} \right) 
    \, \, \text{ for electrons and,} \\
  &\psi_{h-} (\epsilon, \vec{r}\,) = e^{i \vec{p}_{h} \cdot \vec{r}} 
  \left( \begin{array}{c} 0 \\ 0 \\ \frac{p_{x h} - i q}{\mathfrak{p}_{h}} \\ 1 \end{array} \right) 
    \, \, \text{, } \,\,
  \psi_{h+} (\epsilon, \vec{r}\,) = e^{i \vec{p}_{h} \cdot \vec{r}} 
  \left( \begin{array}{c} 0 \\ 0 \\ -\frac{p_{x h} - i q}{\mathfrak{p}_{h}} \\ 1 \end{array} \right) 
    \, \, \text{ for holes.}
\end{align} 
For a given energy ($\epsilon$) and transverse wave vector ($q$), the longitudinal wave vector ($p_{x}$) 
may assume two values: $p_{x} = \pm k_{x}$ where
\begin{align}
  k_{x e} &= \sqrt{(E_{F} + \epsilon_{e})^2 - q^2} \,\, \text{ for electrons and,} \\
  k_{x h} &= \sqrt{(E_{F} - \epsilon_{h})^2 - q^2} \,\, \text{ for holes.}
\end{align} 
These solutions may be propagating ($p_{x}$ is real) or evanescent ($p_{x}$ is imaginary) for 
$|q| \leq \mathfrak{p}$ or $|q| > \mathfrak{p}$. \\

\noindent
The eigenfunctions above may be normalized such that they support unit current along the $x$ direction. 
The absolute value of the average current along $x$ in the states above is
\begin{align}
  &|\langle J_{x} \rangle| = 2 \frac{k_{x}}{\mathfrak{p}} \,\, \text{ for propagating solutions, and} \\
  &|\langle J_{x} \rangle| = 0 \,\, \text{ for evanescent solutions.}
\end{align}
For propagating solutions the eigenfunctions depend only on the angle $\tan^{-1} \frac{q}{p_{x}}$, and not on 
value of $E_{F}$ or $\mathfrak{p}$. \\

\noindent
\underline{{\it Electron-doping -- }}
In the case of electron doping ($E_{F} > 0$), the low energy ($\epsilon \ll E_{F}$) solutions are 
$\epsilon_{e+}$ (with $\mathfrak{p}_{e} > E_{F}$) and $\epsilon_{h-}$ (with $\mathfrak{p}_{h} < E_{F}$). 
For electrons (holes), the states with $p_{x} > 0$ propagate along positive (negative) $x$ direction and
may be labelled as right-movers (left-movers). The opposite is true for states with $p_{x} < 0$.  \\

\noindent
\underline{{\it Hole-doping -- }}
In the case of hole doping ($E_{F} < 0$), the low energy ($\epsilon \ll |E_{F}|$) solutions are 
$\epsilon_{e-}$ (with $\mathfrak{p}_{e} < |E_{F}|$) and $\epsilon_{h+}$ (with $\mathfrak{p}_{h} > |E_{F}|$). 
Now for electrons (holes), the states with $p_{x} > 0$ propagate along negative (positive) $x$ direction and
may be labelled as left-movers (right-movers). The opposite is true for states with $p_{x} < 0$.  \\

\noindent
\underline{{\it Propagating Modes -- }}
For real $p_{x}$ and $q$, we may define a phase $e^{-i \theta_{e/h}} = (p_{x e/h} - iq)/\mathfrak{p}_{e/h}$, 
where $\theta_{e/h} \in (-\pi, \pi]$ is the argument of $\vec{p}$. The eigenfunctions above may obviously be
expressed in terms of $e^{-i \theta_{e/h}}$. For later convenience, we label the eigenfunctions in terms of 
the direction of motion. 
We define $\alpha_{e/h} \in [-\pi/2, \pi/2]$ to be the relative angle between direction of motion and $x$-axis. 
Note that here we define $\alpha$ such that,  $\alpha_{e/h} > 0$ for $q > 0$. 
We also assume that $k_{e}, k_{h} > 0$. \\ 

\noindent
For $E_{F} > 0$, $\alpha_{e} = \theta_{e}$ for right-moving electrons ($p_{x} > 0$) and $\alpha_{e} = \pi - \theta_{e}$ 
for left-moving electrons ($p_{x} < 0$). Similarly, $\alpha_{h} = \pi - \theta_{h}$ for right-moving holes ($p_{x} < 0$) 
and $\alpha_{h} = \theta_{h}$ for left-moving holes ($p_{x} > 0$). Then, after normalizing for current along $x$, we have,
\begin{align}
  \psi_{e, R} (\epsilon, \vec{r}\,) &= \frac{e^{i (k_e x + q y)}}{\sqrt{\cos \alpha_{e}}} \left( \begin{array}{c}
  e^{-i \alpha_{e}} \\ 1 \\ 0 \\ 0 \end{array} \right) 
  \text{, } \,\, 
  \psi_{e, L} (\epsilon, \vec{r}\,) = \frac{e^{i (-k_e x + q y)}}{\sqrt{\cos \alpha_{e}}} \left( \begin{array}{c}
  - e^{i \alpha_{e}} \\ 1 \\ 0 \\ 0 \end{array} \right) 
  \,\, \text{, } \\ \,\, 
  \psi_{h, R} (\epsilon, \vec{r}\,) &= \frac{e^{i (-k_h x + q y)}}{\sqrt{\cos \alpha_{h}}} \left( \begin{array}{c}
  0 \\ 0 \\ -e^{i \alpha_{h}} \\ 1 \end{array} \right) 
  \text{, } \,\, 
  \psi_{h, L} (\epsilon, \vec{r}\,) = \frac{e^{i (k_h x + q y)}}{\sqrt{\cos \alpha_{h}}} \left( \begin{array}{c}
  0 \\ 0 \\ e^{-i \alpha_{h}} \\ 1 \end{array} \right) .
\end{align} \\

\noindent
Similarly for the hole-doped case ($E_{F} < 0$), we may define, $\alpha_{e} = \pi - \theta_{e}$ for 
right-moving electrons ($p_{x} < 0$) and $\alpha_{e} = \theta_{e}$ for left-moving electrons 
($p_{x} > 0$), and $\alpha_{h} = \theta_{h}$ for right-moving holes ($p_{x} > 0$) and 
$\alpha_{h} = \pi - \theta_{h}$ for left-moving holes ($p_{x} < 0$). Then we have, 
\begin{align}
  \psi_{e, R} (\epsilon, \vec{r}\,) &= \frac{e^{i (-k_e x + q y)}}{\sqrt{\cos \alpha_{e}}} \left( \begin{array}{c}
  e^{i \alpha_{e}} \\ 1 \\ 0 \\ 0 \end{array} \right) 
  \text{, } \,\, 
  \psi_{e, L} (\epsilon, \vec{r}\,) = \frac{e^{i (k_e x + q y)}}{\sqrt{\cos \alpha_{e}}} \left( \begin{array}{c}
  - e^{-i \alpha_{e}} \\ 1 \\ 0 \\ 0 \end{array} \right) 
  \,\, \text{, } \\ \,\, 
  \psi_{h, R} (\epsilon, \vec{r}\,) &= \frac{e^{i (k_h x + q y)}}{\sqrt{\cos \alpha_{h}}} \left( \begin{array}{c}
  0 \\ 0 \\ -e^{-i \alpha_{h}} \\ 1 \end{array} \right) 
  \text{, } \,\, 
  \psi_{h, L} (\epsilon, \vec{r}\,) = \frac{e^{i (-k_h x + q y)}}{\sqrt{\cos \alpha_{h}}} \left( \begin{array}{c}
  0 \\ 0 \\ e^{i \alpha_{h}} \\ 1 \end{array} \right) .
\end{align}

\subsection*{States in the Superconducting Region}

\noindent
Graphene with induced superconductivity is described by,
\begin{align}
  H_S = \left( \begin{array}{cc}
    H_0 - E_{F}^{\prime} & \Delta e^{i \phi} \\
  \Delta e^{-i \phi} & E_{F}^{\prime} - H_0 \end{array} \right)
\end{align}
where, $E_{F}^{\prime}$ is the Fermi energy in the superconducting region.
$E_{F}^{\prime}$ is positive (negative) in electron-doped (hole-doped) superconductors. 
In what follows, we shall use the notation $\nu_{E} = \text{sign}(E_{F}^{\prime})$. \\

\noindent
$H_{S}$ has four eigenvalues for each wave vector $\vec{p}$. The two positive eigenvalues are
$\epsilon_{\pm} = \sqrt{\Delta^{2} + (E_{F}^{\prime} \pm \sqrt{p_{x}^{2} + q^{2}})^{2}}$. 
For electron-doping (hole-doping), the low energy branch is $\epsilon_{-}$ ($\epsilon_{+}$). 
The high-energy branch is relevant for $\epsilon \geq |E_{F}^{\prime}|$ and will be ignored here. 
Thus this analysis may need modifications if the Fermi level is tuned close to the SDP. \\ 

\noindent
Since the spectrum has a gap (equal to $\Delta$), it supports both propagating ($\epsilon > \Delta$) and 
evanescent modes ($\epsilon < \Delta$). 
In both cases, $p_{x}$ may assume one of four allowed values for each $\epsilon$ and $q$. 
For the low energy branch, these are,
\begin{align} \label{eq:px1}
  &p_{x} = \pm \sqrt{\big[ E_{F}^{\prime 2} - q^{2} - (\Delta^{2} - \epsilon^{2})\big] 
    \pm 2i |E_{F}^{\prime}| \sqrt{\Delta^{2} - \epsilon^{2}}}  \,\,\,\,\, \text{ for } \epsilon < \Delta \\  \label{eq:px2}
  &p_{x} = \pm \sqrt{\big[ E_{F}^{\prime 2} - q^{2} + (\epsilon^{2} - \Delta^{2})\big] 
    \pm 2 |E_{F}^{\prime}| \sqrt{\epsilon^{2} - \Delta^{2}}}  \,\,\,\,\, \text{ for } \epsilon > \Delta
\end{align}
For brevity, we shall label the 4 solutions as $p_{x; \nu_{o} \nu_{i}}$ where $\nu_{o,i} = \pm$ denote the two choices of sign that appear 
outside ($\nu_{o}$) and inside ($\nu_{i}$) the square root in (\ref{eq:px1},\ref{eq:px2}) above. \\ 

\noindent
To find the corresponding eigenvectors we write the eigenvalue equation after fixing the global phase,
\begin{align}
  \left( 
  \begin{array}{cccc}
    -E_{F}^{\prime} & p_x - i q & \Delta^{i\phi} & 0 \\
     p_x + i q & -E_{F}^{\prime} & 0 & \Delta^{i\phi} \\
    \Delta^{-i\phi} & 0 & E_{F}^{\prime} & - p_x + i q \\
     0 & \Delta^{-i\phi} & -p_x - i q & E_{F}^{\prime} 
  \end{array} \right) \left(
  \begin{array}{c}
    a \\ b \\ e^{-i \phi} \\ d e^{-i \phi} 
  \end{array} \right) = \epsilon \left(
  \begin{array}{c}
    a \\ b \\ e^{-i \phi} \\ d e^{-i \phi} 
  \end{array} \right)
\end{align}
Solving the equations from top 3 rows we find, 
\begin{align}
  a &= \frac{-2 \Delta E_{F}^{\prime}}{p_{x}^{2} + q^2  + \Delta^{2} - (E_{F}^{\prime} + \epsilon)^{2}} \\
  b &= \frac{(E_{F}^{\prime} + \epsilon) a - \Delta}{p_{x} - i q} \\ 
  d &= \frac{(E_{F}^{\prime} - \epsilon) + \Delta a}{p_{x} - i q} 
\end{align}
The solution above holds for all $E_{F}^{\prime}$ and both $\epsilon > \Delta$ and $\epsilon < \Delta$. 
In the latter case, $p_x$ becomes complex and the Hamiltonian is represented by a non-Hermitian matrix,
but the eigenvector above remains valid. \\ 

\noindent
\underline{{\it Evanescent Modes -- }}
For modes with $\epsilon < \Delta$, we use (\ref{eq:px1}) to solve for $p_{x}$ at a given $\epsilon$ and $q$. 
Plugging this solution in the eigenfunction,  we find, 
\begin{align}
  a_{\nu_{o} \nu_{i}} &= \frac{-2 \Delta E_{F}^{\prime}}{p_{x; \nu_{o} \nu_{i}}^{2} + q^2  + \Delta^{2} - (E_{F}^{\prime} + \epsilon)^{2}} 
  =  \frac{\Delta}{\epsilon - i \nu_{i} \nu_{E} \sqrt{\Delta^{2} - \epsilon^{2}}} = e^{i \nu_{i} \nu_{E} \beta}, \\
  b_{\nu_{o} \nu_{i}} &= \frac{(E_{F}^{\prime} + \epsilon) a_{\nu_{o} \nu_{i}} - \Delta}{p_{x; \nu_{o} \nu_{i}} - i q} = 
  d_{\nu_{o} \nu_{i}} a_{\nu_{o} \nu_{i}} = d_{\nu_{o} \nu_{i}} e^{i \nu_{i} \nu_{E} \beta}, \\
  d_{\nu_{o} \nu_{i}} &= \frac{(E_{F}^{\prime} - \epsilon) + \Delta a_{\nu_{o} \nu_{i}}}{p_{x; \nu_{o} \nu_{i}} - i q}  = 
  \frac{E_{F}^{\prime} + i \nu_{i} \nu_{E} \sqrt{\Delta^{2} - \epsilon^{2}}}{p_{x; \nu_{o} \nu_{i}} - i q}.
\end{align} 
where we defined $\beta = \cos^{-1} \big( \frac{\epsilon}{\Delta} \big)$ and $\nu_{E} = \text{sign}(E_{F}^{\prime})$. 
Therefore, for each energy and $q$ we have four wavefunctions of the form, 
\begin{align}
  \psi_{\nu_{o} \nu_{i}} (\epsilon, \vec{r}\,) &= 
  e^{i (p_{x; \nu_{o} \nu_{i}} x + q y)} \left( 
  \begin{array}{c}
    e^{i \nu_{i} \nu_{E} \beta} \\ 
    d_{\nu_{o} \nu_{i}} e^{i \nu_{i} \nu_{E} \beta} \\ e^{-i \phi} \\ d_{\nu_{o} \nu_{i}} e^{-i \phi}
  \end{array} \right)  
\end{align}

\noindent
\underline{{\it Propagating Modes -- }}
For modes with $\epsilon > \Delta$, we use (\ref{eq:px2}) to solve for $p_{x}$ at a given $\epsilon$ and $q$. 
Plugging this solution in the eigenfunction,  we find, 
\begin{align}
  a_{\nu_{o} \nu_{i}} &= \frac{-2 \Delta E_{F}^{\prime}}{p_{x; \nu_{o} \nu_{i}}^{2} + q^2  + \Delta^{2} - (E_{F}^{\prime} + \epsilon)^{2}} 
  =  \frac{\Delta}{\epsilon - \nu_{i} \nu_{E} \sqrt{\epsilon^{2} - \Delta^{2}}} = e^{\nu_{i} \nu_{E} \beta}, \\
  b_{\nu_{o} \nu_{i}} &= \frac{(E_{F}^{\prime} + \epsilon) a_{\nu_{o} \nu_{i}} - \Delta}{p_{x; \nu_{o} \nu_{i}} - i q} = 
  d_{\nu_{o} \nu_{i}} a_{\nu_{o} \nu_{i}} = d_{\nu_{o} \nu_{i}} e^{\nu_{i} \nu_{E} \beta}, \\
  d_{\nu_{o} \nu_{i}} &= \frac{(E_{F}^{\prime} - \epsilon) + \Delta a_{\nu_{o} \nu_{i}}}{p_{x; \nu_{o} \nu_{i}} - i q}  = 
  \frac{E_{F}^{\prime} + \nu_{i} \nu_{E} \sqrt{\epsilon^{2} - \Delta^{2}}}{p_{x;\nu_{o} \nu_{i}} - i q}.
\end{align} 
where we defined $\beta = \cosh^{-1} \big( \frac{\epsilon}{\Delta} \big)$ and $\nu_{E} = \text{sign}(E_{F}^{\prime})$. 
Therefore, for each energy and $q$ we have four wavefunctions of the form, 
\begin{align}
  \psi_{\nu_{o} \nu_{i}} (\epsilon, \vec{r}\,) &= 
  e^{i (p_{x; \nu_{o} \nu_{i}} x + q y)} \left( 
  \begin{array}{c}
    e^{\nu_{i} \nu_{E} \beta} \\ 
    d_{\nu_{o} \nu_{i}} e^{\nu_{i} \nu_{E} \beta} \\ e^{-i \phi} \\ d_{\nu_{o} \nu_{i}} e^{-i \phi}
  \end{array} \right)  
\end{align} \\

\subsection*{Conductance across an N--S--N Junction}

\noindent
Consider a planar N--S--N junction in which the length of the S region (along $x$) is $L_{s}$. 
The setup is assumed to be uniform along the $y$ direction. 
Suppose the injected particle is a right-moving electron in the left lead with energy $\epsilon$. 
Then the wavefunctions in the three regions must be of the form, 
\begin{align}
  &\Psi_{N,\text{Left}} (\epsilon, \vec{r}\,) = \psi_{e,R} (\epsilon, \vec{r}\,)  
    + r_{ee} \psi_{e,L} (\epsilon, \vec{r}\,)+ r_{eh} \psi_{h, L} (\epsilon, \vec{r}\,) , \\ 
  &\Psi_{N,\text{Right}} (\epsilon, \vec{r}\,) = t_{ee} \psi_{e,R} (\epsilon, \vec{r}\,) + t_{eh} \psi_{h,R} (\epsilon, \vec{r}\,) \,\, \text{ and} \\
  &\Psi_{S} (\epsilon, \vec{r}\,) = \sum_{\nu_{o},\nu_{i}} m_{\nu_{o},\nu_{i}} \psi_{\nu_{o},\nu_{i}} (\epsilon, \vec{r}\,). 
\end{align}
Here $r$ and $t$ are the reflection and transmission amplitudes. These wavefunctions describe electrons within the same layer of graphene, and 
hence these should be continuous at the two interfaces, so that, 
\begin{align}
  \Psi_{N,\text{Left}} (x = 0) &= \Psi_{S} (x = 0) , \\ \Psi_{S} (x = L_{s}) &= \Psi_{N,\text{Right}} (x = L_{s}). 
\end{align}
These equations can be partially solved by eliminating the $m$'s. To do this, we rewrite $\Psi_{S}$ as, 
\begin{align}
  &\Psi_{S} (\epsilon, \vec{r}\,) = 
  \mathcal{U}_{S} (\epsilon, \vec{r}\,) \left( \begin{array}{c} 
    m_{++} \\ m_{-+} \\ m_{+-} \\ m_{--}
  \end{array} \right) .
\end{align}
Here, $\mathcal{U}_{S} (\epsilon, \vec{r}\,)$ is a $4 \times 4$ matrix whose columns are the eigenvectors of the 4 states in $S$, 
\begin{align}
  &\mathcal{U}_{S} (\epsilon, \vec{r}\,) 
  = \Big( \begin{array}{cccc} 
    \psi_{++} (\epsilon, \vec{r}\,) & 
    \psi_{-+} (\epsilon, \vec{r}\,) & 
    \psi_{+-} (\epsilon, \vec{r}\,) & 
    \psi_{--} (\epsilon, \vec{r}\,) 
  \end{array} \Big). 
\end{align}
Then we may write the continuity equations as, 
\begin{align} \label{eq:Tmat}
  \Psi_{N,\text{Right}} (x = L_{s}) &= \mathcal{T}(\epsilon) \Psi_{N,\text{Left}} (x = 0), \,\, \text{where } \,\, 
  \mathcal{T}(\epsilon) = \mathcal{U}_{S}(\epsilon, x = L_{s}) \mathcal{U}_{S}(\epsilon, x = 0)^{-1}. 
\end{align}
The matrix $\mathcal{U}_{S}(\epsilon, x = 0)$ is invertible at all parameters, except for $\epsilon = \Delta$ (i.e. $\beta = 0$) and 
$\epsilon = \sqrt{(E_{F}^{\prime})^{2} + \Delta^{2}}$ (for which two of the $d_{\nu_{o} \nu_{i}} = 0$).  
Hence, this procedure is well defined in the range of interest, except for some isolated points.  
Since $\mathcal{T}$ only depends on the eigenvectors in the S region, we may be able to evaluate it analytically in certain limits 
for which the eigenvectors of S assume a simple form. In general it may be computed numerically. \\

\noindent
The $\mathcal{T}$ matrix, which relates the wavefunctions on the left and right of S, may be easily used to find the relation
between the amplitudes of the eigenvectors on the two sides. The general form of the wavefunction in normal regions may be written as, 
\begin{align}
  &\Psi_{N} (\epsilon, \vec{r}\,) = 
  \mathcal{U}_{N} (\epsilon) \left( \begin{array}{c} 
    a_{eR} (\vec{r}\,)\\ a_{eL} (\vec{r}\,)\\ a_{hR} (\vec{r}\,)\\ a_{hL}(\vec{r}\,)
  \end{array} \right) \,\, \text{ where } \,\, 
  \mathcal{U}_{N} (\epsilon) 
  = \Big( \begin{array}{cccc} 
    \psi_{eR} (\epsilon) & 
    \psi_{eL} (\epsilon) & 
    \psi_{hR} (\epsilon) & 
    \psi_{hL} (\epsilon) 
  \end{array} \Big). 
\end{align}
Note that, unlike the case of $\mathcal{U}_{S}$, here we have absorbed the propagation phase $e^{i \vec{p} \cdot \vec{r}}$ in 
the amplitudes $a$. Then $\mathcal{U}_{N}$ is completely independent of the coordinates. Plugging this in (\ref{eq:Tmat}) we find, 
\begin{align}
\left( \begin{array}{c} 
  a_{eR} (x = L_{s})\\ a_{eL} (x = L_{s})\\ a_{hR} (x = L_{s})\\ a_{hL}(x = L_{s})
\end{array} \right)_{\text{Right N}} = \tilde{\mathcal{T}}(\epsilon) 
\left( \begin{array}{c} 
    a_{eR} (x = 0)\\ a_{eL} (x = 0)\\ a_{hR} (x = 0)\\ a_{hL}(x = 0)
\end{array} \right)_{\text{Left N}} \,\, \text{ where } \,\, 
\tilde{\mathcal{T}}(\epsilon)  = \Big[  \mathcal{U}_{N} (\epsilon) \Big]^{-1} \mathcal{T}(\epsilon)  \mathcal{U}_{N} (\epsilon) 
\end{align}
For the problem considered above, we find, 
\begin{align}
\left( \begin{array}{c} 
  t_{ee} e^{i p_{e} L_{s}} \\ 0 \\ t_{eh} e^{i p_{h} L_{s}}\\ 0 
\end{array} \right)_{\text{Right N}} = \tilde{\mathcal{T}}(\epsilon) 
\left( \begin{array}{c} 
  1\\ r_{ee}\\ 0\\ r_{eh}
\end{array} \right)_{\text{Left N}}  
\end{align}
Here, $p_{e/h}$ may be $+k_{e/h}$ or $-k_{e/h}$ depending on the sign of $E_{F}$. However, crucially, we may extract 
two equations for just the reflection coefficients through the second and fourth rows of the matrix equation above. 
This may be rewritten as, 
\begin{align}
\left( \begin{array}{c} 
  r_{ee} \\ r_{eh} 
\end{array} \right) = - \left( \begin{array}{cc} 
  \tilde{\mathcal{T}}_{22} & \tilde{\mathcal{T}}_{24} \\
  \tilde{\mathcal{T}}_{42} & \tilde{\mathcal{T}}_{44} 
\end{array} \right)^{-1} \left( \begin{array}{c}
  \tilde{\mathcal{T}}_{21}\\ \tilde{\mathcal{T}}_{41}
\end{array} \right) . 
\end{align}
Thus we may find the reflection coefficients without worrying about the phase factors appearing with the $t$'s. \\

\noindent
The differential conductance of the setup may be computed through the Landauer formalism, as applied to the superconducting case by BTK, 
\begin{align}
  g_{NP} = \frac{4e^{2}}{h} \sum_{q} 1 - |r_{ee}|^{2} + |r_{eh}|^{2}
\end{align}
where the factor of 4 accounts for spin and valley degeneracies.  
Suppose the width of the setup (along $y$) is $W$, then we have, 
\begin{align}
  \sum_{q} = \frac{W}{2\pi} \int_{-E_{F}-\epsilon}^{E_{F}+\epsilon} \, dq 
  = \frac{W (E_{F} + \epsilon)}{2 \pi} \int_{-\pi/2}^{\pi/2} \, d\theta \, \cos \theta
\end{align}
so that the differential conductance is, 
\begin{align}
  g_{NSN} &= \frac{8e^{2}}{h} \frac{W (E_{F} + \epsilon)}{2 \pi} \int_{0}^{\pi/2} \, d\alpha \, \cos \alpha \, 
  \Big[ 1 - |r_{ee}|^{2} + |r_{eh}|^2 \Big] \\
  &= g_{0} \int_{0}^{\pi/2} \, d\alpha \, \cos \alpha \, \Big[ 1 - |r_{ee}|^{2} + |r_{eh}|^2 \Big], 
\end{align}
where $g_{0}$ accounts for the density of states, and the integral accounts for the effect of the interface.


\subsection*{Analytical Result at Normal Incidence} 

\noindent
Consider the special case of normal incidence ($q = 0$). Then we find the wavevectors in N and S regions to be, 
\begin{align}
  k_{xe} &= |E_{F} + \epsilon| \,\, \text{ and } k_{xh} = |E_{F} - \epsilon|, \\ 
  p_{x; \nu_{o} \nu_{i}} &= \nu_{o} \Big[ |E_{F}^{\prime}| 
  + i \nu_{i} \Delta \sin \beta \Big] \,\, \text{ for } \epsilon < \Delta, \\
  p_{x; \nu_{o} \nu_{i}} &= \nu_{o} \Big[ |E_{F}^{\prime}| 
  + \nu_{i} \Delta \sinh \beta \Big] \,\, \text{ for } \epsilon > \Delta. 
\end{align}
Using these in the definition of $d$, we find, 
\begin{align}
  d_{\nu_{o} \nu_{i}} &= \nu_{o} \nu_{E} \,\, \forall \,\, \epsilon. 
\end{align}
Thus the eigenfunctions are particularly simple in this limit, and we may find the conductance analytically. 
In what follows, we only focus on the regime of $\epsilon > \Delta$. The results for energies below the gap can be 
obtained by replacing $\beta$ by $i \beta$. 
Then the $\mathcal{U}_{S}$ matrices are, 
\begin{align}
  \mathcal{U}_{S}( \epsilon, x = y = 0)^{-1} &= 
  \frac{1}{4 \sinh \beta}\left(
\begin{array}{cccc}
  1 & 1 & - e^{-\beta} & - e^{-\beta} \\
 1 & -1 & - e^{-\beta} & e^{-\beta } \\
 -1 & -1 & e^{\beta } & e^{\beta } \\
 -1 & 1 & e^{\beta } & -e^{\beta } 
\end{array}
  \right) \,\, \text{ for } \epsilon > \Delta, \\
  \mathcal{U}_{S}( \epsilon, x = L_{s}, y = 0) &= 
  \left( \begin{array}{cccc}
    e^{i p_{++} L_{s}} e^{\beta } & e^{\beta}  e^{-i p_{++} L_{s}} & e^{i p_{+-}L_{s}} e^{-\beta } &  e^{-i p_{+-}L_{s}} e^{ -\beta } \\
    e^{i p_{++} L_{s}} e^{\beta } & -e^{\beta} e^{-i p_{++} L_{s}} & e^{i p_{+-}L_{s}} e^{-\beta } & -e^{-i p_{+-}L_{s}} e^{ -\beta } \\
    e^{i p_{++} L_{s}}            &  e^{-i p_{++} L_{s}} & e^{i p_{+-} L_{s}} &  e^{-i p_{+-} L_{s}} \\
    e^{i p_{++} L_{s}}            & -e^{-i p_{++} L_{s}} & e^{i p_{+-} L_{s}} & -e^{-i p_{+-} L_{s}} \\
\end{array} \right) \,\, \text{ for } \epsilon > \Delta 
\end{align}
And after some algebra, the transfer matrix $\mathcal{T}$ may be written down in terms of $2 \times 2$ blocks, 
\begin{align}
  \mathcal{T} (\epsilon) &=   \frac{1}{ 2 \sinh (\beta) } 
  \left(
  \begin{array}{cc}
    \mathcal{T}_{1} &
    \mathcal{T}_{2} \\
    \mathcal{T}_{3} &
    \mathcal{T}_{4} 
  \end{array} \right), \,\, \text{ where the blocks are}
\end{align}
\begin{align}
  \mathcal{T}_{1} &= 
\left(
\begin{array}{cc}
  e^{\beta } \cos \left(p_{++} L_s\right)-e^{-\beta } \cos \left(p_{+-} L_s\right) & 
  i \left(e^{\beta} \sin \left(p_{++} L_s\right) - e^{-\beta} \sin \left(p_{+-} L_s\right)\right)  \\
  i \left(e^{\beta} \sin \left(p_{++} L_s\right) - e^{-\beta} \sin \left(p_{+-} L_s\right)\right)  & 
  e^{\beta } \cos \left(p_{++} L_s\right)-e^{-\beta } \cos \left(p_{+-} L_s\right)  
\end{array}
\right) , \\
  \mathcal{T}_{2} &= -\mathcal{T}_{3} = 
  \left(
\begin{array}{cc}
  \cos \left(p_{+-} L_s\right)-\cos \left(p_{++} L_s\right) & 
  i \sin \left(p_{+-} L_s\right) - i \sin \left(p_{++} L_s\right) \\
  i \sin \left(p_{+-} L_s\right) - i \sin \left(p_{++} L_s\right) & 
  \cos \left(p_{+-} L_s\right)-\cos \left(p_{++} L_s\right) 
\end{array}
\right) , \\ 
  \mathcal{T}_{4} &= 
  \left(
\begin{array}{cc}
  e^{\beta } \cos \left(p_{+-} L_s\right)-e^{-\beta } \cos \left(p_{++} L_s\right) & 
  i e^{\beta } \sin \left(p_{+-} L_s\right) - i e^{-\beta } \sin \left(p_{++} L_s\right) \\
  i e^{\beta } \sin \left(p_{+-} L_s\right) - i e^{-\beta } \sin \left(p_{++} L_s\right) & 
  e^{\beta } \cos \left(p _{+-} L_s\right)-e^{-\beta } \cos \left(p_{++} L_s\right) 
\end{array}
\right). 
\end{align}
The transfer matrix in terms of amplitudes $\tilde{\mathcal{T}}$ is, 
\begin{align}
  \tilde{\mathcal{T}} &= \frac{1}{\sinh \beta}
  \left(
\begin{array}{cccc}
  e^{i |E_{F}^{\prime}| L_s} \tilde{t}_{1} & 0 & 0 & 
  -i e^{i |E_{F}^{\prime}| L_s} \tilde{t}_{2} \\
 0 & 
  e^{-i |E_{F}^{\prime}| L_s} \tilde{t}_{1}^{*} & 
  i e^{-i |E_{F}^{\prime}| L_s} \tilde{t}_{2} & 0 \\
 0 & -i e^{-i |E_{F}^{\prime}| L_s} \tilde{t}_{2} & 
  e^{-i |E_{F}^{\prime}| L_s} \tilde{t}_{1} & 0 \\
 i e^{i |E_{F}^{\prime}| L_s} \tilde{t}_{2} & 0 & 0 & 
  e^{i |E_{F}^{\prime}| L_s} \tilde{t}_{1}^{*} 
\end{array}
  \right) , \,\, \text{ where} \\
  \tilde{t}_{1} &= \sinh \left(\beta + i L_s \Delta \sinh \beta\right) , \\
  \tilde{t}_{2} &= \sin \left(L_s \Delta \sinh \beta\right). 
\end{align}
Finally, this can be used find the reflection amplitudes. For normal incidence we find, 
\begin{align}
  r_{ee} &= 0 , \\
  r_{eh} &= -i \frac{\sin \left( L_{s} \Delta \sinh \beta \right)}
  {\cos \left( L_{s} \Delta \sinh \beta \right) \sinh \beta - i \sin \left( L_{s} \Delta \sinh \beta \right) \cosh \beta}. 
\end{align}
Therefore the transmission probability through a superconducting segment at normal incidence is, 
\begin{align}
  1 - |r_{ee}(q = 0)|^{2} + |r_{eh}(q = 0)|^{2} &= 
  1+\frac{1 - \cos\left(2 L_{s} \Delta \sinh \beta \right)}{\cosh (2 \beta ) - \cos \left(2 L_s \Delta \sinh \beta  \right)}. 
\end{align}
Clearly, the transmission probability oscillates with increasing $\beta$ (or energy $\epsilon$), and has a minimum 
for all $\beta$ satisfying, 
\begin{align}
  L_{s} \Delta \sinh \beta = n \pi  \,\, \text{ for } n = 1, 2, 3, \dots 
\end{align}
Since the total conductance is dominated by the transmission probability close to normal incidence, we expect minima in the differential conductance at these values of the bias above the energy gap. Putting back all units, we find the bias for the minima to be at, 
\begin{align}
  \epsilon_{n} =  \sqrt{\Delta^2 + \left( n \frac{ \pi\hbar v_{F}}{L_{s}} \right)^{2}}. 
\end{align}


\subsection*{Analytical Approximations for the General Case}
To conclude, we perform an approximate calculation of $r_{ee},r_{eh}$ in the case of $\alpha\neq 0$. We start by parameterizing:

\begin{align}
    \alpha = & \sin^{-1} \left(\frac{q}{E_F}\right) \\
    \beta = & \cos^{-1} \left(\frac{\epsilon}{\Delta}\right) \\
    \delta \pm \zeta = & \sin^{-1}\left(\frac{q}{\left|E_F^\prime\pm \sqrt{\epsilon^2-\Delta^2}\right|}\right) \\
    \theta = & \sqrt{E_F^{\prime2}-q^2}L \\
    \eta = & \frac{\left|E_F^\prime\right| \sqrt{\epsilon^2-\Delta^2}}{\sqrt{E_F^{\prime2}-q^2}}L \\
    \nu_E = & sign(E_F^\prime) = \frac{E_F^\prime}{\left|E_F^\prime\right|}
\end{align}

Here $\alpha$ is the incidence angle; $\beta$ is a parametrization of $\epsilon$ (note that this differs from the definition in the main paper by a factor of $i$); $\delta\pm\zeta$ are the two transmission angles inside the superconducting regime; and $\theta\pm \eta$ are the phases gatheres while going through the SC regime, namely $k_x L$. $\eta$ is the phase splitting, of order $L\Delta$ (for $\epsilon$ of order $\Delta$).

We approximate $\zeta\approx 0$ as, in the relevant regime of parameters, $\epsilon,\Delta \ll E_F^\prime$. 
Following the calculations described in the previous parts, we can calculate the elements of $\tilde{\mathcal{T}}$ and extract $r_{ee}, r_{eh}$. We define the angular quantities

\begin{align}
    \tau = & \sin(\alpha)\sin(\delta) - \nu_E \\
    \sigma = & \sin(\delta) - \nu_E \sin(\alpha)\\
    \chi = & \cos(\delta)\cos(\alpha)
\end{align}

and obtain the reflection amplitude

\begin{align}
    r_{ee} = \frac{i \sigma e^{-i\alpha} \sin(\nu_E \beta)(\chi(\cos\eta \sin \eta \cos(\nu_E\beta)+ i \sin\theta \cos \theta \sin(\nu_E\beta)) + \tau \sin(\nu_E\beta) (\sin^2\theta - \sin^2\eta))}{\chi^2(\sin(\eta+\nu_E \beta))^2 + \frac{1}{2} \chi(\tau-\chi) \sin(2\eta)\sin(2\nu_E \beta)+ \sigma^2\sin(\nu_E \beta)(\sin^2(\theta)-\sin^2(\eta))} \\
    r_{eh} = \frac{-\chi \sin(\eta) (\tau \cos(\eta) \sin(\nu_E\beta) + \chi \sin(\eta) \cos(\nu_E\beta))}{\chi^2(\sin(\eta+\nu_E \beta))^2 + \frac{1}{2} \chi(\tau-\chi) \sin(2\eta)\sin(2\nu_E \beta)+ \sigma^2\sin(\nu_E \beta)(\sin^2(\theta)-\sin^2(\eta))}\,.
\end{align}
This expression is too complicated to integrate over, but we can draw some simple conclusions - for example, we see how $\eta$ and $\theta$ cause oscillations as a function of $L$. We can also substitute $\alpha=0$, hence $q=0$ and $\delta=0$. The angular quantities are $\tau = -\nu_E$, $\sigma=0$ and $\chi=1$, immediately leading to $r_{ee}=0$ as $r_{ee} \propto \sigma$. For the Andreev reflection, we obtain the expression

\begin{align}
    r_{eh} = & \frac{-\sin(\eta) (-\cos(\eta) \sin(\beta) + \sin(\eta) \cos(\beta))}{(\sin(\eta+\beta))^2 - \sin(2\eta)\sin(2\beta)} \\
    = & -\frac{\sin(\eta) \sin(\eta-\beta)}{\sin(\eta-\beta)^2} = -\frac{\sin(\eta)}{\sin(\eta-\beta)}\,;
\end{align}
substituting $\beta \mapsto i\beta$ (this is more useful for $\epsilon > \Delta$, which is the case of interest), we use $|\sin(\eta-i\beta)|^2=\frac{\cosh(2\beta)-\cos(2\eta)}{2}$ to get

\begin{align}
    |r_{eh}|^2 = \frac{1-\cos(2\eta)}{\cosh(2\beta)-\cos(2\eta)}
\end{align}
which, after substituting $\eta=\frac{\left|E_F^\prime\right| \sqrt{\epsilon^2-\Delta^2}}{\sqrt{E_F^{\prime2}-q^2}}L=\sqrt{\epsilon^2-\Delta^2}L=L\Delta\sinh\beta$, gives the results used in the main text.

Another limit that is analytically achievable is the low-$\Delta$ limit, in which the Andreev reflection vanishes, and the normal reflection amplitude is given by

\begin{align}
    r_{ee}=-e^{-i\alpha} \frac{\sin(\alpha)-\sin(\delta)}{\cos(\delta)\cos(\alpha)\cot(\theta)-i(1-\sin(\delta)\sin(\alpha))}
\end{align}
here we can more clearly see the Klein paradox (for $\alpha=\delta=0$, $r=0$) and also the $\theta=k_xL=\sqrt{E_F^{\prime2}-q^2}$ dependence.

\end{document}